\documentclass[conference]{IEEEtran}
\IEEEoverridecommandlockouts
\usepackage{cite}
\usepackage{amsmath,amssymb,amsfonts}
\usepackage{algorithmic}
\usepackage{graphicx}
\usepackage{textcomp}
\usepackage{xcolor}
\def\BibTeX{{\rm B\kern-.05em{\sc i\kern-.025em b}\kern-.08em
    T\kern-.1667em\lower.7ex\hbox{E}\kern-.125emX}}
\begin{document}

\title{Single Image Super-Resolution Methods: A Survey\\}

\author{\IEEEauthorblockN{Bahattin Can MARAL}
\IEEEauthorblockA{\textit{Department of Computer Engineering} \\
\textit{TOBB University of Economics and Technology}\\
Ankara, Turkey \\
bahattincanmaral@etu.edu.tr} \\
}

\maketitle

\begin{abstract}
Super-resolution (SR), the process of obtaining high-resolution images from one or more low-resolution observations of the same scene, has been a very popular topic of research in the last few decades in both signal processing and image processing areas. Due to the recent developments in Convolutional Neural Networks\cite{lawrence1997face}, the popularity of SR algorithms has skyrocketed as the barrier of entry has been lowered significantly. Recently, this popularity has spread into video processing areas to the lengths of developing SR models that work in real-time. In this paper, we compare different SR models that specialize in single image processing and will take a glance at how they evolved to take on many different objectives and shapes over the years. 
\end{abstract}

\section{Introduction}
Image Super-Resolution (SR) is the process of achieving high-detailed, high-resolution (HR) images from one or multiple low-resolution (LR) observations of the same scene. Rapid developments in image processing and deployment of scene recognition for visual communications have created a strong need for high-resolution images not only to provide better visualization (fidelity) but also for the extraction of additional information details (recognition). High-resolution images are useful when isolating regions in multi-spectral remote sensing images \cite{zhang2020scene,courtrai2020small,gong2021enlighten,arun2020cnn,zhang2020remote} or when they assist radiologists in making diagnostic decisions based on the images \cite{isaac2015super,greenspan2009super,kouame2009super,christensen2020super,gupta2020super,gu2020medsrgan,he2020super}. When it comes to video surveillance systems, higher-resolution video frames are always appreciated for more accurate identification of the objects and people of interest. In order to obtain higher-resolution images, the most direct means is to reduce the pixel size on the sensor (e.g., charge-coupled device) of an image acquisition device (e.g., digital camera); sensor technology, however, has limitations when it comes to reducing sensor pixels. The quality of captured images will inevitably deteriorate if the sensor's pixel size is too small, signal power decreases proportionally to the reduction in pixel size, while noise power remains roughly the same. In addition, a larger chip incurs a higher cost. SR image processing is, therefore, an attractive alternative because of the factors listed.

Despite having been explored for decades, image super-resolution remains a challenging task in computer vision. The ill-posed nature of this problem stems from the fact that each LR image can contain multiple HR images that have slight differences in the camera angle, color, brightness, and other variables. Moreover, the LR and HR data are subject to fundamental uncertainties, since it is possible to downscale two HR images to yield the same LR image. In a nutshell, it is a many-to-one conversion.

The methods of image super-resolution available today are either single-image super-resolution (SISR) or multiple-image methods. When using single-image SR, each LR-HR pair within the image is learned separately, while in multiple-image SR, the LR-HR pairs within a scene are learned to be able to generate an HR image from the scene (multiple images). In video super-resolution, multiple successive images (frames) are super-resolved using the relationship between them; it is a special form of multiple image SR, defined as an image that is part of a scene comprised of different frames.

Traditional methods of achieving super-resolution in the past include statistical methods, prediction-based methods, patch-based methods, and edge-based methods. Recently, the advance in computational power and big data has prompted researchers to use deep learning (DL)\cite{dong2014learning} to address the problem of SR. SR studies based on deep learning have featured superior performance than classical methods in the past decade, and DL methods are commonly used to achieve SR. A variety of methods have been used to investigate SR, from the first Convolutional Neural Network (CNN)\cite{lawrence1997face} to the latest Generative Adversarial Nets (GAN)\cite{ledig2017photo}. There is no consensus on the hyperparameters that make up deep learning-based SR methods, such as network architecture, learning strategies, activation functions, and loss functions.

In this study, a brief overview of the classical methods of SR is outlined initially, whereas the main focus is given to give an overview of the most recent research in SR using deep learning, specifically on SISR.

\section{Early Days of Super-Resolution}
Image Super-resolution (SR) techniques try to construct a high resolution (HR) image from one or more observed low resolution (LR) images\cite{yue2016image}. Due to SR's ill-posed nature, many possible solutions exist. Concerning LR input images, SR techniques can be divided into two main groups, namely single-image super-resolution (SISR) and multiple-image super-resolution or multi-frame super-resolution. As the SISR requires only one input LR image to produce a corresponding HR image, it has attracted the attention of researchers as it is closer to everyday life settings. 

Early SISR techniques can be divided into two types:
\begin{enumerate}
 
\item \textbf{The learning-based methods}. These types of SISR methods use machine-learning techniques to estimate HR images. Pixel-based methods\cite{zhang2012single} and example-based methods\cite{freeman2002example} are typical methods in this category. Other techniques like sparse coding and neighbor embedding are also widely used\cite{gao2012image}. 

\item \textbf{The reconstruction-based methods}. Which needs prior knowledge to define constraints for the target HR image. Typically, techniques like edge sharpening\cite{dai2007soft}, regularization\cite{aly2005image}, and deconvolution\cite{shan2008fast} are employed in this category.

\end{enumerate}
The problem of SISR has been alleviated by using methods to learn prior information from LR and HR pairs. Examples include neighbor embedding regression \cite{timofte2013anchored}, random forests \cite{schulter2015fast}, and deep convolutional neural networks\cite{dong2014learning}. 

\section{Deep Learning Era of Image Super-Resolution}
Computer vision applications have become more robust with deep learning \cite{lecun2015deep}, especially convolutional neural networks (CNNs) \cite{krizhevsky2012imagenet}. Although CNNs aren't perfect \cite{shalev2017failures}, their performance in different computer vision applications has been reported to be outstanding \cite{zeiler2014visualizing,szegedy2015going}. This section discusses recent SISR methods based on CNNs and relative methods.

\subsection{Convolutional Neural Networks}
\begin{figure*}

  \includegraphics[width=\textwidth]{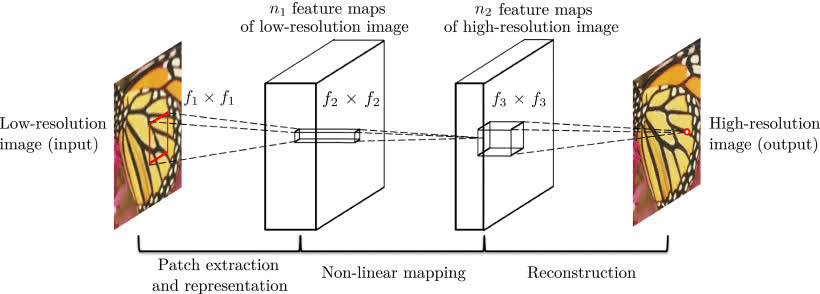}
  \caption{SRCNN network structure \cite{dong2015image}}
  \label{fig:srcnn}
\end{figure*}

SRCNN \cite{dong2015image} is the first CNN-based SISR model, illustrated in Figure \ref{fig:srcnn}. In the method, the input LR image is mapped to the HR image by learning the end-to-end mapping. This technique employs bicubic interpolation as a pre-processing step. After that, it extracts feature vectors from the image patches by convolution, which are then non-linearly mapped to find the most representative patches to reconstruct the HR image. SRCNN only uses convolutional layers, so it's possible to input images of any size, and its algorithm is not patch-based \cite{kappeler2016video}. The SRCNN model outperforms many "traditional" models. 

\begin{figure*}
  \includegraphics[width=\linewidth]{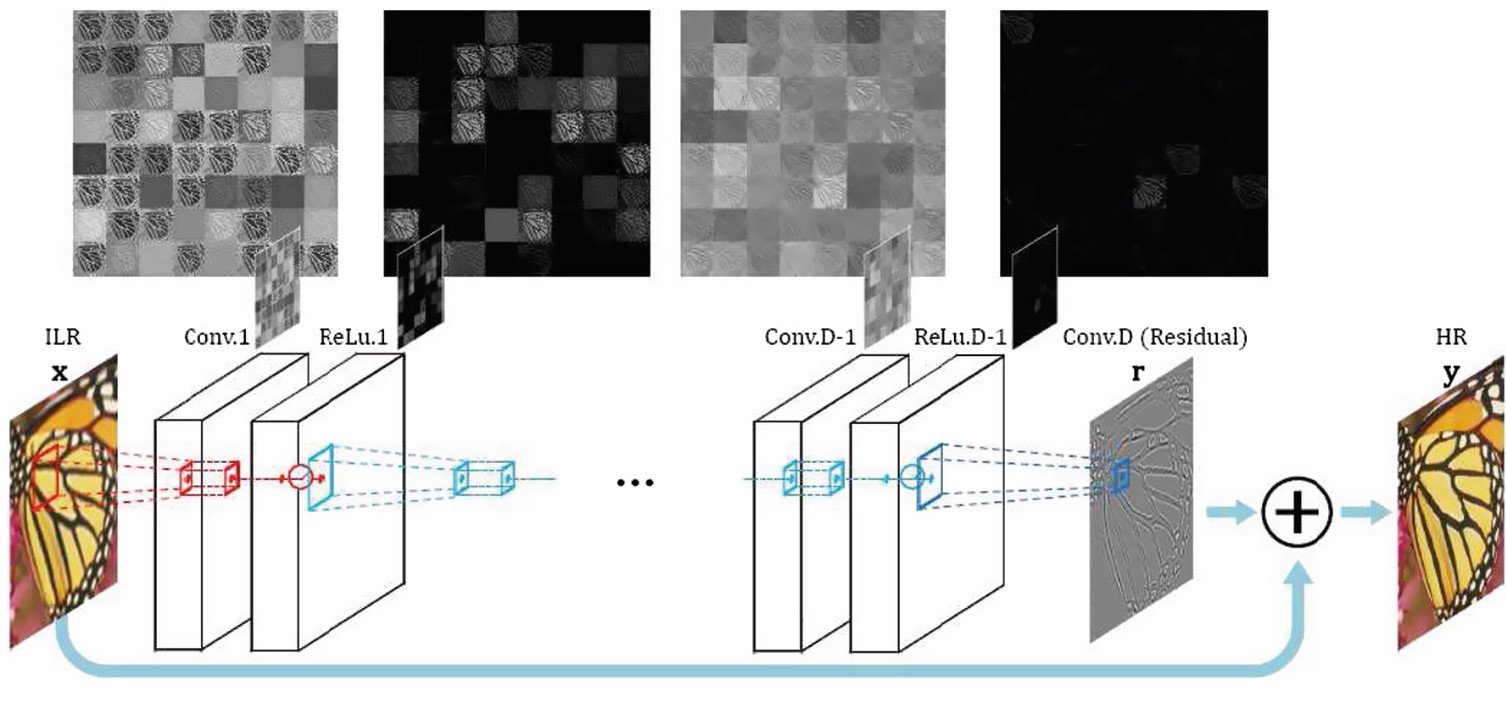}
  \caption{VDSR \& DRCN network structure \cite{kim2016accurate}}
  \label{fig:vdsr}
\end{figure*}

Based on this simple model, it appears possible that the accuracy cannot be further improved.  As a result, the question arose whether "the deeper, the better" holds in SR or not. Following the success of very deep networks, Kim went on to propose two new algorithms called Very Deep Convolutional Networks (VDSR)\cite{kim2016accurate} and Deeply Recursive Convolutional Networks (DRCN)\cite{kim2016deeply}, which both used 20 convolutional layers, as illustrated in Figure \ref{fig:vdsr}.  Gradient clipping was used to control the explosion problem, while the VDSR was trained with a very high learning rate to accelerate the convergence speed.

Currently, the peaks of “vanilla” SISR CNN performance and accuracy are dominated by the neural network architecture type called u-Net\cite{ronneberger2015u}, which uses staggered “leaky” layers to compute the HR image. This type of neural network architecture is highly compressible and lightweight compared to the other CNN architecture types that usually use stacked dense layers.

\begin{figure*}
  \includegraphics[width=\linewidth]{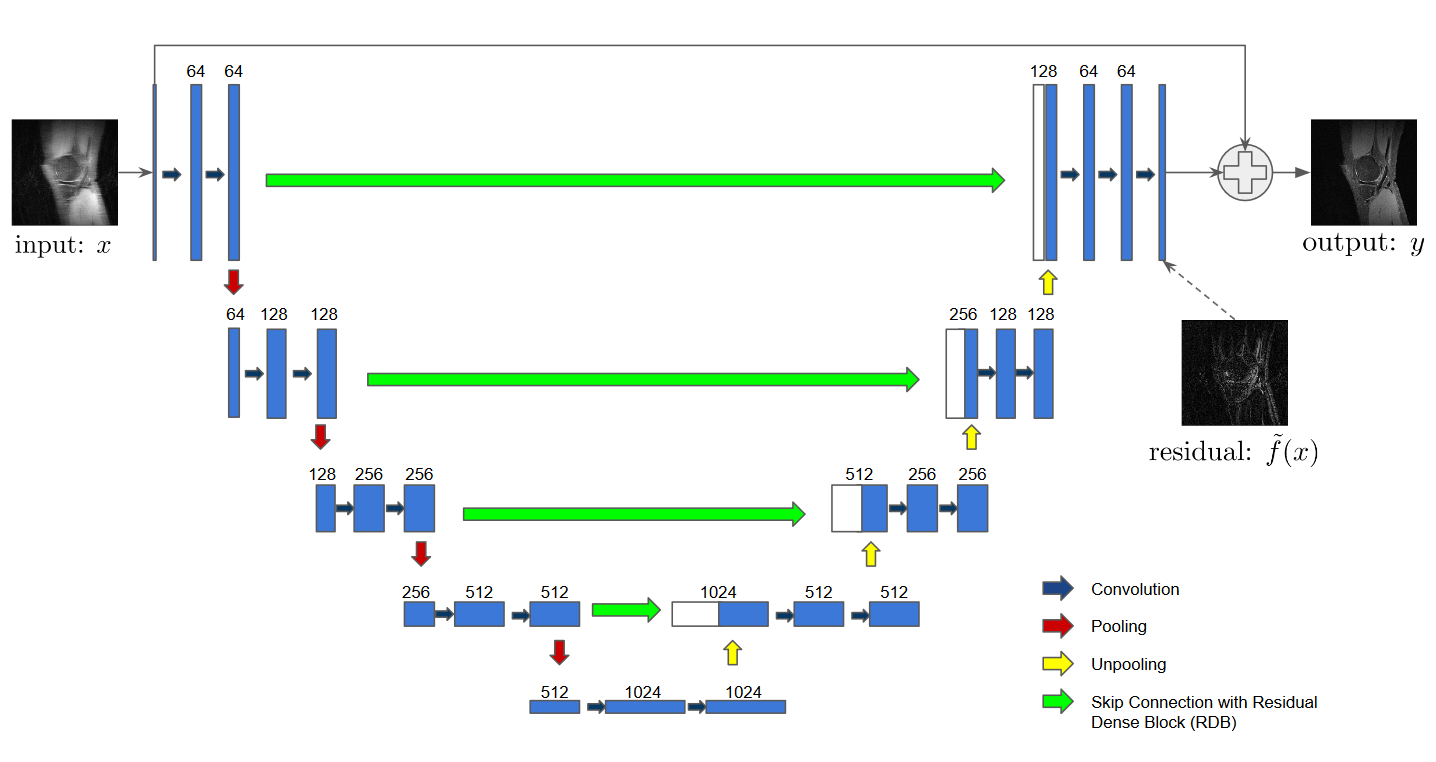}
  \caption{UDSR Network architecture \cite{ding2019deep}}
  \label{fig:superrior}
\end{figure*}

The world’s most competitive image processing workshop NTIRE (New Trends in Image Restoration and Enhancement) annually organizes various machine learning challenges (image restoration, downscaling, recoloring, etc.). The most recent SISR challenge\cite{cai2019ntire} was won by the SuperRior team that was utilizing a u-net architecture. This model is referred to as U-shaped Deep Super-Resolution (UDSR) and an illustration of it is shown in Figure \ref{fig:superrior}. A convolution layer is used in UDSR to extract deep feature maps from a low-resolution input image. The feature maps were then processed by residual blocks and down-sampled to a lower resolution. In order to obtain high-resolution feature maps, they upsampled the feature maps, as well as applied more residual blocks. The left side of the U-shaped network was connected to the right side by straight paths. To create the final output, they used a residual image that was derived from the highest resolution feature maps. 

In addition, they train models by using a cascaded approach, in order to refine the input image better with each stage. A UDSR is used to process the input image from each stage, using the output of the previous stage as input. Three stages have different supervision signals, from coarse to fine. They first downsample the high-resolution ground truth by four scales and then upsample it to the original size. The 4× blurred image is used to supervise the output of the first UDSR model. Second, the output of the second UDSR model is supervised using the blurred 2x image. As part of the third stage, the ground-truth image is used to supervise the UDSR model of the third stage.

Finally, the results were merged using an ensemble of adaptive multi-models. Diverse models have different characteristics. Even with the same model, the performance of different patches varies a great deal. Moreover, these priors motivate them to ensemble multiple models in an adaptive way, namely, the fusion weights of different models must be conditioned on the frames generated by these models in an image patch granularity. They use a CNN model to operate the outputs of several models and learn a normalized weight for each pixel of every single model.

Another u-net inspired model can be found in \cite{9166725}, in which the authors mainly focused on the efficiency and the lightweightness of the model to use them in computers with less computing power (like in IoTs). The lightweightness comes from the less dense layers of neurons.  

\subsection{Adaptive Models}
In addition to using CNNs for classification tasks, many researchers build SISR models that are more adaptive to the content of images (pixels or structures). \cite{liang2017single} presents a Deep Projection CNN (DPN) method. Model adaptation in DPN is used to seek out repetitive structures in LR images. \cite{dahl2017pixel} proposes the pixel recursive super-resolution network, which consists of a conditioning network and a prior network. Conditional networks transform LR images into logits, resulting in multiple predictions of the likelihood of each HR pixel. Prior networks are called pixelCNNs\cite{oord2016conditional}. Models built in this way can add realistic details to images and enhance resolution at the same time. The authors of \cite{wang2015self} propose a model named deep joint super-resolution (DJSR) in order to adapt the deep model for joint similarities. 

In 2018, researchers of \cite{zhang2018adaptive} proposed an adaptive residual network (ARN) for high-quality image restoration. The ARN, which consists of six cascaded adaptive shortcuts, convolutional layers, and PReLUs, is a deep residual network. Each adaptive shortcut contains two small convolutional layers, followed by PReLU activation layers and one adaptive skip connection. It is possible to train the ARN model depending on the application.

\begin{figure*}
  \includegraphics[width=\linewidth]{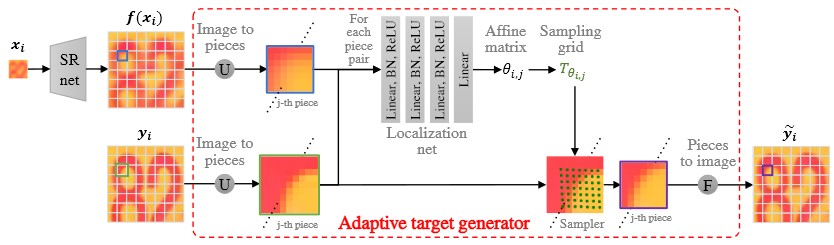}
  \caption{Network structure of the adaptive target generator \cite{Jo_2021_CVPR}}
  \label{fig:atg}
\end{figure*}

One of the most recent examples of adaptive models utilizes adaptive models for target generation. In \cite{Jo_2021_CVPR}, the authors describe a simple and effective way to cultivate sharp output generation by accepting solutions other than those provided by the training pair. The new method calculates the loss based on an adaptive target yi instead of directly comparing to the original target yi. In theory, their alternative target allows different HR predictions based on LR input to relax the typical pixel reconstruction loss. Adaptive targets are made from the original targets so that the network prediction f(xi) is penalized at the lowest rate, while maintaining the original contents and perceptual impression. In particular, they find an affine transform matrix for every small non-overlapping piece of yi to those of f(xi) within the range of acceptable transforms. After that, each piece is transformed to construct the adaptive target, Figure \ref{fig:atg}. This can be done on-the-fly during training with minimal computational overhead. During each training iteration, the SR network is trained using the loss computed with the adapted target.
\subsection{Generative Adversarial Network Based Models}
In contrast to traditional machine learning methods, generative adversarial networks (GANs) are known for their ability to preserve texture details in images, create solutions that are close to the real image, and appear perceptually convincing. Thus, GANs are also suitable for SISR. The authors in \cite{mao2016super} propose the Depixelated Super-Resolution Convolutional Neural Network (DSRCNN). It is designed to resolve partially pixelated images for super-resolution. Depixelation is achieved by combining hnbp autoencoder with depixelating layers. The autoencoder is composed of a generator and a discriminator. In \cite{bosch2018super}, a GAN-based architecture using densely connected convolutional neural networks (DenseNets) is proposed for super resolving overhead imagery by as much as 8×.

In \cite{ledig2017photo}, the most known and first successful GAN-based SISR model, the Super-Resolution Generative Adversarial Network (SRGAN) is introduced, of which a generative network upsamples LR images to super-resolution (SR) images, and the discriminative network is to distinguish the ground truth HR images and SR images. The pixel-wise quality assessment metric has been criticized for showing poor human perception. With the addition of adversarial loss, GAN-based algorithms were able to improve perceptive naturalistic images. By fusing pixel-wise loss, perceptual loss, and newly proposed texture transfer loss, the GAN-based SISR model has been further developed in \cite{demiray2021d,wang2018esrgan}. The SRFeat proposed by Park et al.\cite{park2018srfeat} employed an additional discriminator in the feature domain. There are two phases to training the generator: pre-training and adversarial training. During the pre-training phase, the generator is trained to minimize MSE losses to achieve high PSNR.  By using perceptual similarity loss, GAN loss in the pixel domain, and GAN loss in the feature domain, the training procedure is aimed at improving perceptual quality. A major disadvantage of GAN-based SISR methods is the difficulty of training the models.

\subsection{Sparsity Based Models}
Researchers have shown that sparse coding combined with CNNs can produce better performance than CNNs alone \cite{ding2018deeply,shi2017structure,osendorfer2014image,wang2015deep}. Using sparse priors, the model of the sparse coding-based network (SCN) in \cite{liu2016robust} is more compact and accurate than the SRCNN. During the training of a deep CNN, another model, SCRNN-Pr \cite{liang2016incorporating}, explores image priors as well. Compared with other current state-of-the-art methods, better training time cost and super-resolution tasks are reported. 
\cite{gao2016hybrid} proposes a hybrid wavelet convolution network (HWCN). LR images are fed into a scattering convolution network (a wavelet tree in nature) to obtain scattering feature maps. Sparse codes are then extracted from these maps and used to input a CNN. With this model, complex deep networks can be trained with a tiny dataset with better generalization. 

In \cite{wang2015self}, a sparse representation-based noise-robust super-resolution approach that incorporates smoothing prior to enforcing the same sparse coding coefficients in similar training patches is proposed. It employs LASSO-based smooth constraint combined with locality-based smooth constraint for obtaining stable reconstruction weights, especially when noise levels are high in the input LR image.

\begin{figure}
  \includegraphics[width=\linewidth]{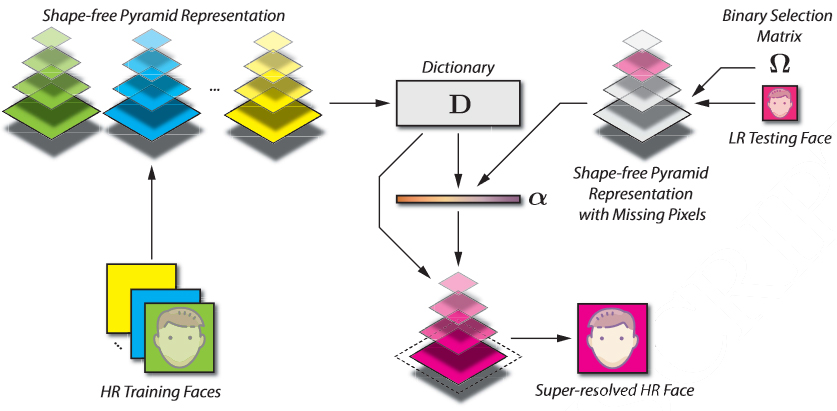}
  \caption{Overview of the SSR2 method. \cite{ABIANTUN2019308}}
  \label{fig:ssr2}
\end{figure}

In one of the most recent examples of sparsity-based SISR research papers\cite{ABIANTUN2019308}, the SISR model named SSR2 is specifically tuned to extract and amplify information from extreme low resolution images of human faces. The overall similarity in human faces represent a base output for the model to build on top of it. With this logic, sparsity is used to extract only the most defining features from the lower resolution face image and amplified to get a 16x super-resolution. The base human face is also helpful to get rid of the obstructions that may occur in the low resolution image. The overall method of SSR2 can be seen at Figure \ref{fig:ssr2}. This model is made to enhance human faces to recognizable sizes from surveillance camera footage.
\section{Discussion}
In this paper, we decided to focus on the different types of Single Image Super-Resolution models rather than focusing on individual works done by other researchers. The reason behind this is that while under the same broad term of SISR they are structured and optimized for entirely different tasks. In our observations, we deduced that a SISR model that has been built to work on LR images that have a lower amount of information than others tend to use sparsity-based SISR structures. The SISR models that have been built for extreme super-resolution tasks can be given as an example for this category.
On the other hand, the GAN-based SISR models dominated the research areas that worked on images with similar attributes, like human faces, medical imagery, etc. This can be attributed to the tear-down build-from-scratch nature of the GAN-based models.
We also observed that the adaptive models while not the most popular in the SISR world did wonder on the restoration of details and LR images with subpar qualities.
CNN-based models were dominating the areas where the desired SR images didn’t have many details like animation images, space photography, and alike. 

GAN-based models were by far the most popular way of implementing SISR, this comes from the ease of implementation and decent performance in every type of research area.

When it comes to the best possible performance in any type of objective image, we see a combination of the model types listed. The models with this method of learning are called ensemble models. Ensemble models are at the peak of any given objectives because they make up for the downfalls of the main model by introducing side models with different types.

Examples of ensemble learning can be found in \cite{liu2020ensemble,liu2020progressive,cheng2020self,shang2020perceptual,zhang2020pqa,shahsavari2021proposing,lyu2020mri,pan2020real}, and it’s easy to see why are they better at their jobs than their vanilla competitors. Ensemble models while finicky to set up and train, they’ve proved their worth time and time again with their performance.

\section{Conclusion}

In this paper, we took a closer look at why the different types of SISR models are still trained for different types of image categories, and what they are best at. 
In conclusion, we observed that there is no best type of structure when it comes to the ill-posed SISR problem. The objective of your project and the type of images you use greatly impact the models’ performance. That’s why there are predominant structures when it comes to different research topics. 
Lately, ensemble models are commonly praised and used in the most revolutionary research papers of any given topic.

In the close future that has no scientific breakthroughs in deep learning, we expect SISR models to become more specialized for the type of image used, and to be on the structure of ensemble models. 
Currently. for a general-purpose SISR model, mainly GAN-based ensemble models are the way to go. We expect this to change into an ensemble model where there can be a pre-deciding machine learning model that chooses the percentages of the different types of SISR model outputs in the ensemble model itself. This approach could mitigate the downfalls of model types greatly while being adaptable to any image type possible.

\bibliography{references}
\bibliographystyle{IEEEtran}
\end{document}